\documentclass[twocolumn,showpacs,preprintnumbers,amsmath,amssymb,aps]{revtex4}
\usepackage{graphicx}
\usepackage{dcolumn}
\usepackage{bm}
\begin{document}
\title{Dissipative dynamics of a Harmonic Oscillator : A non-perturbative approach}
\author{Jishad Kumar, S. Sinha, P. A. Sreeram}
\affiliation{ Indian Institute of Science Education and Research, Kolkata,700106 }
\date{\today}
\begin{abstract}
Starting from a microscopic theory, we derive a master equation for a harmonic oscillator coupled to a bath of non-interacting oscillators. We follow a non-perturbative approach, proposed earlier by us for the free Brownian particle. The diffusion constants are calculated analytically and the positivity of the Master Equation is shown to hold above a critical temperature. We compare the long time behaviour of the average kinetic and potential energies with known thermodynamic results. In the limit of vainishing oscillator frequency of the system, we recover the results of the free Brownian particle. 
\end{abstract} 
\pacs{To be appear here}
\maketitle
\section{Introduction}

The problem of quantum dissipation has been a long standing and important problem and has been studied in the context of a variety of phenomena, including relaxation phenomena\cite{weiss,sdg}, quantum cosmological  models\cite{zurek,halliwell} and in more recent times, in the context of cold atoms\cite{recati}. In contrast to the classical dissipative phenomenon, the quantum analogue has a number of additional constraints, which on the one hand make the problem much more complex, and on the other hand makes it more interesting. Of these, the first and the foremost constraint is that of the non-commutative nature of the canonical variables, which leads to stringent conditions on the fluctuation of the canonical co-ordinates, which is commonly known as the Heisenberg uncertainity relations. The second major difference between the classical and quantum dissipative dynaimcs, is the fact that the autocorrelation fucntion of the fluctuating force has a memory effect, which depends on the temperature and the Planck Constant $\hbar$. Thus, while the force autocorrelation becomes delta-correlated (Markovian) at high temperatures or $\hbar \rightarrow$ 0, leading to recovery of the classical results, the correlation can have a power law dependence on time at very low temperatures\cite{leggett2}. Hence, a consistent theory has to extrapolate between these two extreme regimes. Another major issue of the quantum dissipative system is the positivity of the density matrix during the non-equilibrium dynamics of the system. The positivity is necessary for the evolution of the system through physical states.

Historically, several approaches have been taken to describe the quantum Brownian Motion \cite{grabert,ingold-chaos}.
Senitzky\cite{senitzky}, applied the quantum dissipation model to radiating electrmagnetic waves in a cavity and showed that even without the details of the dissipative medium being known, one can formulate a mechanism which leads to quantum dissipation.  One of the first attempts in solving the problem was by Zwanzig\cite{zwanzig}, who considered the long range memory effect in the fluctuating force autocorrelation function. Ford, Kac and Mazur\cite{fkm} proposed the system-bath model for the classical and quantum dissipative system, where the system is linearly coupled to a bath of harmonic oscillators. A formal treatment of the system plus bath model, was introduced by Feyman and Vernon\cite{feynman}, who derived the influence functional for the effective action of the system, by integrating out the bath degrees of freedom, using the path integral technique. This method was later used by Caldeira and Leggett\cite{cl} to derive the high temperature behaviour of the master equation for a quantum harmonic oscillator coupled to a bath of harmonic oscillators. Dekker\cite{dekker} proposed a phenomenological master equation where there can be diffusion not only in the position space but also in the momentum space. Dekker and Valsakumar\cite{positivity1} showed that the various diffusion constants must obey a cretain criterion in order to satisfy the positivity condition as shown by Lindblad\cite{lindblad,positivity2}. In the weak coupling limit, Agarwal\cite{gsa} derived some exact results for a dissipative harmonic oscillator. Hu, Paz and Zhang\cite{paz} did an exact treatment of the same problem and found that the diffusion constants are time dependent. In all the previous approaches, one of the crucial assumption was that the system and bath are initially decoupled and they get coupled at time $t=0$. Hakim and Ambegaokar\cite{hakim} showed that the results would be dramatically different if this assumption were not true and that the initial transients can play a crucial role in the evolution of the density matrix. There have been some attempts to derive the master equation, phenomenologically, within Lindblad form\cite{gao}.

In this paper, we derive the master equation of a quantum harmonic oscillator, coupled to a heat bath, within a scheme, proposed earlier by us\cite{ss}. Unlike some of the earlier approaches\cite{cl,diosi}, where the high temperature expansion was used to convert the non-local memory kernel to a local one, we retain the original expression for the memory kernel and Taylor expand the non-local terms in the local time scheme. Thus, our results can be valid, in general, for a wide range of temperatures. The paper is organized as follows. In Section II, we formulate the problem of the Harmonic oscillator and show how one can derive the master equation following our scheme. In Section III, we calculate the diffusion constants and remark on their behaviour at high temperatures. We also calculate the limit in which the positivity condition breaks down. In Section IV, we recover the free particle results exactly, in the limit of vanishing oscillator frequency. We conclude in Sectilon V.

\section{Derivation of the Master Equation for the Harmonic Oscillator Coupled to a Heat Bath} 
We start with a microscopic model for a harmonic oscillator, which is linearly coupled to a bath of oscillators. 
The total Hamiltonian $\mathcal{H}$ of the system consists of three parts, viz., the Hamiltonian for the system $\mathcal{H}_{A}$, the Hamiltonian for the bath $\mathcal{H}_{B}$, and the interaction Hamiltonian $\mathcal{H}_{I}$, which describes the linear coupling of the system to the heat bath. The three parts of the total Hamiltonian $\mathcal{H}$ are defined as :
\begin{subequations}
\begin{align}
\mathcal{H}_{A}&=\frac{p^2}{2M}+\frac{1}{2}M\omega_0^{2}q^{2} \label{ha} \\
\mathcal{H}_{B}&=\sum_{i} \left(\frac{P_{i}^2}{2m_{i}}+\frac{1}{2}m_{i}\omega_{i}^{2}Q_{i}^{2}\right)  \label{hb} \\
\mathcal{H}_{I}&=q\sum_{i}C_{i}Q_{i} 
\label{hi}
\end{align}
\label{hamil}
\end{subequations}
where, $q$ and $p$ are the position and momentum coordinates of the system,with mass $M$, oscillator frequency $\omega_0$, and $Q_{i}$ and $P_{i}$ are the coordinates of the bath oscillators, with mass $m_{i}$ and the frequency $\omega_{i}$.  The coupling constant between the system and the $i^{th}$ bath oscillators is denoted by $C_{i}$.

In order to derive a master equation of the reduced density matrix, we use the Feynman-Vernon procedure to calculate the effective action, after integrating out the bath degrees of freedom. We assume that at $t=0$ both the system and the bath are uncorrelated with each other, so that we can write the total initial density matrix of the whole system as the direct product of the initial density matrices of the system and the bath, respectively. Following the work of  Feynman and Vernon\cite{feynman}, the time evolution of the density matrix is given in terms of the Influence Functional, as,
\begin{equation}
\rho(q_{1},q_{2},t)=\int dq_{1}^{\prime}dq_{2}^{\prime}J(q_{1},q_{2},t;q_{1}^{\prime},q_{2}^{\prime},0)\rho(q_{1}^{\prime},q_{2}^{\prime},0)
\label{time_evol}
\end{equation}
Where the quantity $J(q_{1},q_{2},t;q_{1}^{\prime},q_{2}^{\prime},0)$ propagates the density matrix $\rho(q_{1}^{\prime},q_{2}^{\prime},0)$ from an initial  time to the density matrix $\rho(q_{1},q_{2},t)$ at final time, and is given by,
\begin{equation}
J(q_{1},q_{2},t;q_{1}^{\prime},q_{2}^{\prime},0)=\int
\int{\cal D}q_{1}{\cal D}q_{2} \exp\left(\frac{\imath}{\hbar} S_{\rm eff}[q_1,q_2]\right)
\label{propagator}
\end{equation}
Integrating out the bath degrees of freedom, we obtain the nonlocal effective action for the system\cite{cl}, given by,
\begin{equation}
\frac{\imath}{\hbar}S_{{\rm eff}}=\frac{\imath}{\hbar}
(S_{A}[q_{1}]-S_{A}[q_{2}])+\int_{0}^{t}(\Sigma_{R}+\Sigma_{I})d\tau
\label{seff_new}
\end{equation}
where, $S_{A}$ corresponds to the action of system corresponding to Eq. \ref{ha} and,
\begin{subequations}
\begin{align}
\Sigma_{R} &=  -\frac{1}{\hbar}\int_{0}^{\tau}[q_{-}(\tau)\alpha_{R}
(\tau-s)q_{-}(s)]ds  \label{sigma-R}\\
\Sigma_{I} &=  -\frac{\imath}{\hbar}\int_{0}^{\tau}[q_{-}(\tau)\alpha_{I}(\tau-s)q_{+}(s)]ds \label{sigma-I}
\end{align}
\label{sigmas}
\end{subequations}
here we define,$q_{\pm}=q_{1}\pm q_{2}$ and the respective memory kernels are given by the relations, 
\begin{subequations}
\begin{align}
\alpha_{R}(\tau) &= \sum_{i}\frac{C_{i}^{2}}{2m\omega_{i}}\coth
\left(\frac{\hbar\omega_{i}}{2k_{B}T}\right)\cos(\omega_{i}\tau) \label{alpha-R} \\ 
\alpha_{I}(\tau) &=  -\sum_{i}\frac{C_{i}^{2}}{2m\omega_{i}}\sin(\omega_{i}\tau).\label{alpha-I}
\end{align}
\label{memory-kernel} 
\end{subequations}
The summations in Eq. \ref{memory-kernel} can be replaced by integrals, by defining a density of states $F(\omega)$ of the bath oscillators. The density of states, is defined using the Drude form of the cutoff\cite{weiss}, as, 
\begin{equation}
 F(\omega)=\omega_{c}^{2}/(\omega^{2}+\omega_{c}^{2})
\label{fom}
\end{equation}
and
\begin{equation}
F(\omega)C^{2}(\omega)/(2m\omega^{2})=(2M\gamma/\pi)(\omega_{c}^{2}/(\omega^{2}+\omega_{c}^{2})). 
\label{fom1}
\end{equation}
The memory kernels can now be evaluated as:
\begin{eqnarray}
\alpha_{R}(\tau) & = & M\gamma \omega_{c}^{2}\left[\cot(\chi)\exp(-\omega_{c}\tau)\right.\nonumber \\
& + & \left.\frac{2}{\chi}\sum_{n=1}^{\infty}\frac{n\pi/\chi}{(n\pi/\chi)^{2}-1}\exp\left[-(n\pi/\chi)\omega_{c}\tau\right]\right]
\label{alpha-r-1}
\end{eqnarray}
where, $\chi$ = $\hbar\omega_{c}/2k_{B}T$. The above expression shows clearly that the system has two distinct time scales, one given by $1/\omega_c$ and the other by $\hbar/k_BT$.  We now proceed to use our scheme of using a Taylor series expansion, as was done in an earlier work\cite{ss},
\begin{equation}
q_{\pm}(s)=\sum_{l=0}^{\infty}\frac{q_{\pm}^{(l)}(\tau)}{l!}(s - \tau)^{l}
\label{taylor}
\end{equation}
where $q^{(l)}$ is the $l^{th}$ derivative of the position coordinates of the system with respect to time. Inserting Eq. \ref{alpha-r-1} and Eq. \ref{taylor} into Eq. \ref{sigmas}, neglecting the total derivative terms, we obtain,
\begin{equation}
\Sigma_{R}(\tau) = -\frac{1}{\hbar}\sum_{l=0}^{\infty}\frac{(-1)^{l}}{2l!} \left(q_{-}^{(l)}(\tau)\right)^2\int_{0}^{\tau}
\tilde \tau^{2l}\alpha_{R}(\tilde \tau)d\tilde \tau
\label{sigr}
\end{equation}
For $\tau\gg\hbar/k_{B}T$, neglecting the transient terms, we obtain,
\begin{eqnarray}
&&\Sigma_{R}(\tau) =  -\frac{M \gamma \omega_{c}}{\hbar\chi}\times \nonumber \\ 
 && \left[ 
q_{-}^{2}(\tau)
-2 \sum_{l=1}^{\infty}(q_{-}^{(l)})^{2}\frac{(-1)^{l}}{\omega_{c}^{2l}}
\sum_{m=0}^{l}\left(\frac{\chi}{\pi}\right)^{2m}\zeta(2m)\right] 
\label{sigma-r-gen}
\end{eqnarray}
where $\zeta$ is the Riemann Zeta function, and 
\begin{equation}
\Sigma_{I}(\tau)=-\frac{\imath\gamma M}{\hbar}q_{-}(\tau)\dot{q_{+}}(\tau)
\label{sigma-i-gen}
\end{equation}
where the higher order terms in $\Sigma_I$ are negelected as they fall off as $1/\omega_c$.
Note here, that while Eq. \ref{sigma-r-gen} and Eq. \ref{sigma-i-gen}, are local in nature, the nonlocality is implicitly present due to the infinite summation and the presence of all higher order derivatives of the dynamical variables. We now proceed to find a systematic replacement for the higher order derivatives. In order to do this we resort to the dynamical equations of motion.
The dynamical equation of motion for the harmonic oscillator can be written as,
\begin{equation}
\dot {\bf Q} = {\bf M} {\bf Q}
\label{mat-eq} 
\end{equation}
where, 
\begin{eqnarray}
{\bf Q} &=& \left( \begin{array}{c}
q \\ \dot{q} 
          \end{array} \right)  \nonumber \\
{\bf {\cal M}}&=& \left( \begin{array}{cc}
                   0 & 1/M \\
		   -M \omega_0^2 & -2 \gamma/M
                  \end{array} \right)
\label{qandm}
\end{eqnarray}
The higher order derivatives of the position variable $q$, can be formally obtained from the relation,
\begin{equation}
{\bf Q }^{(l+1)} = {\bf {\cal M}}^l {\bf Q}
\label{powers}
\end{equation}
Using the eigenvalues and eigenfunctions of the matrix ${\bf {\cal M}}$, and performing a similarity transformation, the higher order derivatives of the position variable can be written as, 
\begin{equation}
 q^{(l)}=\frac{(\lambda_{2}\lambda_{1}^{l}-\lambda_{1}\lambda_{2}^{l})}{(\lambda_{2}-\lambda_{1})}q+\frac{(\lambda_{2}^{l}-\lambda_{1}^{l})}{\lambda_{2}-\lambda_{1})}\dot{q}
\label{q-l}
\end{equation}
where,
\begin{equation}
\lambda_{1,2} = -\gamma \pm \sqrt{\gamma^2-\omega_0^2} 
\label{lam12}
\end{equation}
are the eigenvalues of the matrix {\bf {\cal M}}. It is interesting to note that all the higher order derivatives can be written in terms of the canonical coordinates at time $t$. 
Inserting Eq. \ref{q-l} into Eq. \ref{sigma-r-gen}, gives,
\begin{equation}
\Sigma_{R}(\tau)=\frac{-2k_{B}T\gamma M}{\hbar^2}[\alpha q_{-}^2(\tau)+\alpha ^{'}\dot{q}_{-}^{2}(\tau)]
\label{sigr11}
\end{equation}
where
\begin{eqnarray}
 \alpha&=&1+\left(\frac{\lambda_{1} \lambda_{2}}{\lambda_{2}-\lambda_{1}}\right)^2 \left\{ \frac{1}{\lambda_{1}^{2}} \left[\frac{\frac{\lambda_{1}\chi}{\omega_{c}}\coth(\frac{\lambda_{1}\chi}{\omega_{c}})}{1+\frac{\lambda_{1}^{2}}{\omega_{c}^{2}}}-1\right]\right . \nonumber \\
&-&\frac{2}{\lambda_{1} \lambda_{2}}\left[\frac{\frac{\sqrt{\lambda_{1} \lambda_{2}}\chi}{\omega_{c}}\coth(\frac{\sqrt{\lambda_{1}\lambda_{2}}\chi}{\omega_{c}})}{1+\frac{(\sqrt{\lambda_{1}\lambda_{2}})^{2}}{\omega_{c}^{2}}}-1\right] \nonumber \\
&+&\frac{1}{\lambda_{2}^{2}}\left . \left[\frac{\frac{\lambda_{2}\chi}{\omega_{c}}\coth(\frac{\lambda_{2}\chi}{\omega_{c}})}{1+\frac{\lambda_{2}^{2}}{\omega_{c}^{2}}}-1\right]\right\}
\label{al1}
\end{eqnarray}
and 
\begin{eqnarray}
\alpha^{\prime}&=&\frac{1}{(\lambda_{2}-\lambda_{1})^2} \left\{ \left[\frac{\frac{\lambda_{1}\chi}{\omega_{c}}\coth(\frac{\lambda_{1}\chi}{\omega_{c}})}{1+\frac{\lambda_{1}^{2}}{\omega_{c}^{2}}}-1\right] \right . \nonumber \\
&-& 2 \left[\frac{\frac{\sqrt{\lambda_{1} \lambda_{2}}\chi}{\omega_{c}}\coth(\frac{\sqrt{\lambda_{1}\lambda_{2}}\chi}{\omega_{c}})}{1+\frac{(\sqrt{\lambda_{1}\lambda_{2}})^{2}}{\omega_{c}^{2}}}-1\right] \nonumber \\
&+&\left . \left[\frac{\frac{\lambda_{2}\chi}{\omega_{c}}\coth(\frac{\lambda_{2}\chi}{\omega_{c}})}{1+\frac{\lambda_{2}^{2}}{\omega_{c}^{2}}}-1\right] \right\}
\label{alp1}
\end{eqnarray}
In the limit of $\omega_c \rightarrow \infty$, $\alpha$ and $\alpha^\prime$ are given by,
\begin{eqnarray}
 \alpha&=&1+\left(\frac{\lambda_{1} \lambda_{2}}{\lambda_{2}-\lambda_{1}}\right)^2 \left\{ \frac{1}{\lambda_{1}^{2}} \left[\frac{\hbar\lambda_{1}}{2k_B T}\coth(\frac{\hbar\lambda_{1}}{2k_B T})-1\right]\right . \nonumber \\
&-&\frac{2}{\lambda_{1} \lambda_{2}}\left[\frac{\hbar\sqrt{\lambda_{1} \lambda_{2}}}{2k_B T}\coth(\frac{\hbar\sqrt{\lambda_{1}\lambda_{2}}}{2k_B T})-1\right] \nonumber \\
&+&\frac{1}{\lambda_{2}^{2}}\left . \left[\frac{\hbar\lambda_{2}}{2k_B T}\coth(\frac{\hbar\lambda_{2}}{2k_B T})-1\right] \right \}
\label{alpha}
\end{eqnarray}
and
\begin{eqnarray}
  \alpha^\prime&=&\left(\frac{1}{\lambda_{2}-\lambda_{1}}\right)^2 \left \{ \left[\frac{\hbar\lambda_{1}}{2k_B T}\coth(\frac{\hbar\lambda_{1}}{2k_B T})-1\right] \right .\nonumber \\
&-&2  \left[\frac{\hbar\sqrt{\lambda_{1} \lambda_{2}}}{2k_B T}\coth(\frac{\hbar\sqrt{\lambda_{1}\lambda_{2}}}{2k_B T})-1\right] \nonumber \\
&+&\left . \left[\frac{\hbar\lambda_{2}}{2k_B T}\coth(\frac{\hbar\lambda_{2}}{2k_B T})-1\right] \right \}
\label{alpha-prime}
\end{eqnarray}
Thus the effective action is given by
\begin{eqnarray}
\frac{\imath}{\hbar}S_{{\rm eff}}  &=& 
 \frac{\imath}{\hbar}\int_{0}^{t} d\tau \left[\frac{M}{2}\dot{q}_{+}
\dot{q}_{-} -\frac{M\omega_0^2}{2} q_+q_--\gamma M q_{-} \dot{q}_{+} \right] \nonumber \\
&-&\frac{2 k_{B} T \gamma M}{\hbar^{2}} \int_{0}^{t} d\tau \left[ \alpha q_{-}^{2}
+ \alpha^{\prime} \dot{q}_{-}^{2} \right]
\label{eff-action}
\end{eqnarray}
It is worth noting that using our scheme of summing the memory effects over the classical paths, we have converted the non-local action into a local effective action. This is the central result of this paper. 

Once the effective action in a local form has been obtained, viz. Eq. \ref{eff-action}, we follow the prescription given by Caldeira and Leggett\cite{cl}, to derive the master equation. We calculate the change in the reduced density matrix, within a small time interval $\epsilon$, using the functional integral method. Collecting the terms which are of the order of $\epsilon$, and neglecting higher order terms, we obtain the master equation, describing the time evolution of the density matrix $\rho(x,y,t)$  in the following form : 
\begin{eqnarray}
 & &\frac{\partial \rho}{\partial t}=\frac{i\hbar}{2M}\left[\frac{\partial^{2} \rho}{\partial x^{2}}-\frac{\partial^{2} \rho}{\partial y^{2}}\right]-\frac{\imath}{\hbar}\frac{M\omega_0^2}{2}(x^2-y^2)\rho \nonumber \\
&-&\gamma (x-y)\left[\frac{\partial \rho}{\partial x}-\frac{\partial \rho}{\partial y}\right] +\frac{2i}{\hbar}D_{pq}(x-y)\left[\frac{\partial \rho}{\partial x}+\frac{\partial \rho}{\partial y} \right] \nonumber \\
&+&D_{qq}\left(\frac{\partial}{\partial x}+\frac{\partial}{\partial y}\right)^2 \rho -\frac{D_{pp}}{\hbar^2}(x-y)^2 \rho
\label{master-eq}
\end{eqnarray}
where the diffusion constants are given by,
\begin{subequations}
\begin{align}
 D_{pq}&=4kT\gamma^2 \alpha^{\prime} \label{dpq1}\\ 
D_{qq}&= \frac{2kT\gamma\alpha^{\prime}}{M} \label{dqq1}\\ 
D_{pp}&=2kTM\gamma(\alpha+4\gamma^{2}\alpha^{\prime}) \label{dpp1}
\end{align}
\label{diff-const-1}
\end{subequations}
The above form of the master equation is different from the one obtained by Caldeira and Leggett\cite{cl}, due to the fact that our effective action has an extra term depending on $\dot {q}^2$. It is relevant to note that the form of the master equation, which we have obtained from a microscopic model, is similar to that obtained by Dekker\cite{dekker}, from a phenomenological point of view. However, the functional form of the diffusion constants are different from those proposed by Dekker. 

\section{Discussions of results}

In this section, we examine the behaviour of the diffusion constants with respect to temperature as well as on the damping and the oscillator frequency. There are two dimensionless parameters in this problems, viz., $\gamma/\omega_0$ and $k_B T/\hbar \omega_0$. The cutoff frequency $\omega_c$ is considered to be much larger than any of the other scales in the problem. In terms of the dimensionless parameter $\gamma/\omega_0$ there are two regimes : (1) the overdamped regime ($\gamma/\omega_0 > 1$) and the underdamped regime ($\gamma/\omega_0 < 1$), which shows oscillatory behaviour. At high temperatures, the diffusion constants are the same for both the regimes and are given by,
\begin{subequations}
\begin{align}
D_{pp} &= 2 k_B T M \gamma \left( 1+\frac{\hbar^2\gamma^2}{3 k_B^2 T^2}\right) \label{dpp2} \\ 
D_{qq} &= \frac{\hbar^2 \gamma}{6 M k_B T} \label{dqq2} \\ 
D_{pq} &= \frac{\hbar^2\gamma^2}{3 k_B T} \label{dpq2}
\end{align}
\label{diff-ht}
\end{subequations}
The high temperature behaviour is completely independent of the oscillator frequency, upto the order of $1/T$. In the high temperature limit, the classical diffusive behaviour is recovered, with the well known diffusion constant, $D_{pp}=2 M \gamma k_B T$, and the anomalous diffusion constants $D_{qq}$ and $D_{pq}$ vanish as $1/T$. 

The two central issues in the problem of quantum brownian motion as discussed earlier are : (1) the Heisenberg Uncertainity Principle and (2) the positivity of the density matrix during the time evolution. Dekker and Valsakumar\cite{positivity1} addressed the first issue and have shown that the diffusion constants have to satisfy a certain condition, known as the positivity condition, in order to maintain the uncertainity relation. The second issue has been addressed by Lindblad\cite{lindblad} from a more mathematical approach, where a general form of the master equation was proposed, in order to maintain the positivity of the density matrix for all times. Interestingly, if the first condition is satisfied, then it automatically guarantees the Dekker equation in a Lindblad form\cite{positivity2}.  In terms of the diffusion constants the Dekker-Valsakumar positivity condition is given by\cite{positivity1}:
\begin{equation}
 \Delta = D_{pp} D_{qq} - D_{pq}^{2} -\hbar^2 \gamma^2/4 > 0
\label{positivity-eq}
\end{equation}
At high temperatures, although the anomalous diffusion constants vanish as $1/T$, they are crucial for restoring the positivity condition and $\Delta$ approaches a constant value $\hbar^2 \gamma^2/12$. Using Eq. \ref{diff-const-1} in Eq. \ref{positivity-eq}, we find that the positivity condition is only satisfied above a breakdown temperature $T_c$ which depends on $\gamma/\omega_0$. 
\begin{figure}[htb]
\vspace{0.5cm}
\begin{center}
\includegraphics[width=7cm]{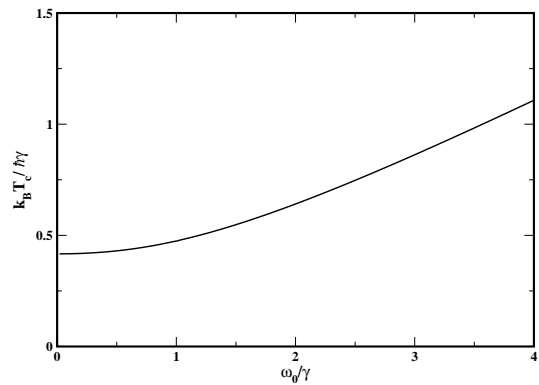}
\end{center}
\caption{The breakdown temperature $T_c$ as a function of the dimensionless parameter $\omega_0/\gamma$.}
\label{critical}
\end{figure}
In Fig. \ref{critical} we plot the dimensionless quantity $k_B T_c/\hbar \gamma$ as a function of the dimensionless parameter $\omega_0/\gamma$. For small $\omega_0$, this critical temperature is of the order of $0.4 \hbar \gamma/k_B$, which agrees with the results for the free particle\cite{ss}.  . 
\begin{figure}[h]
\includegraphics[width=8cm]{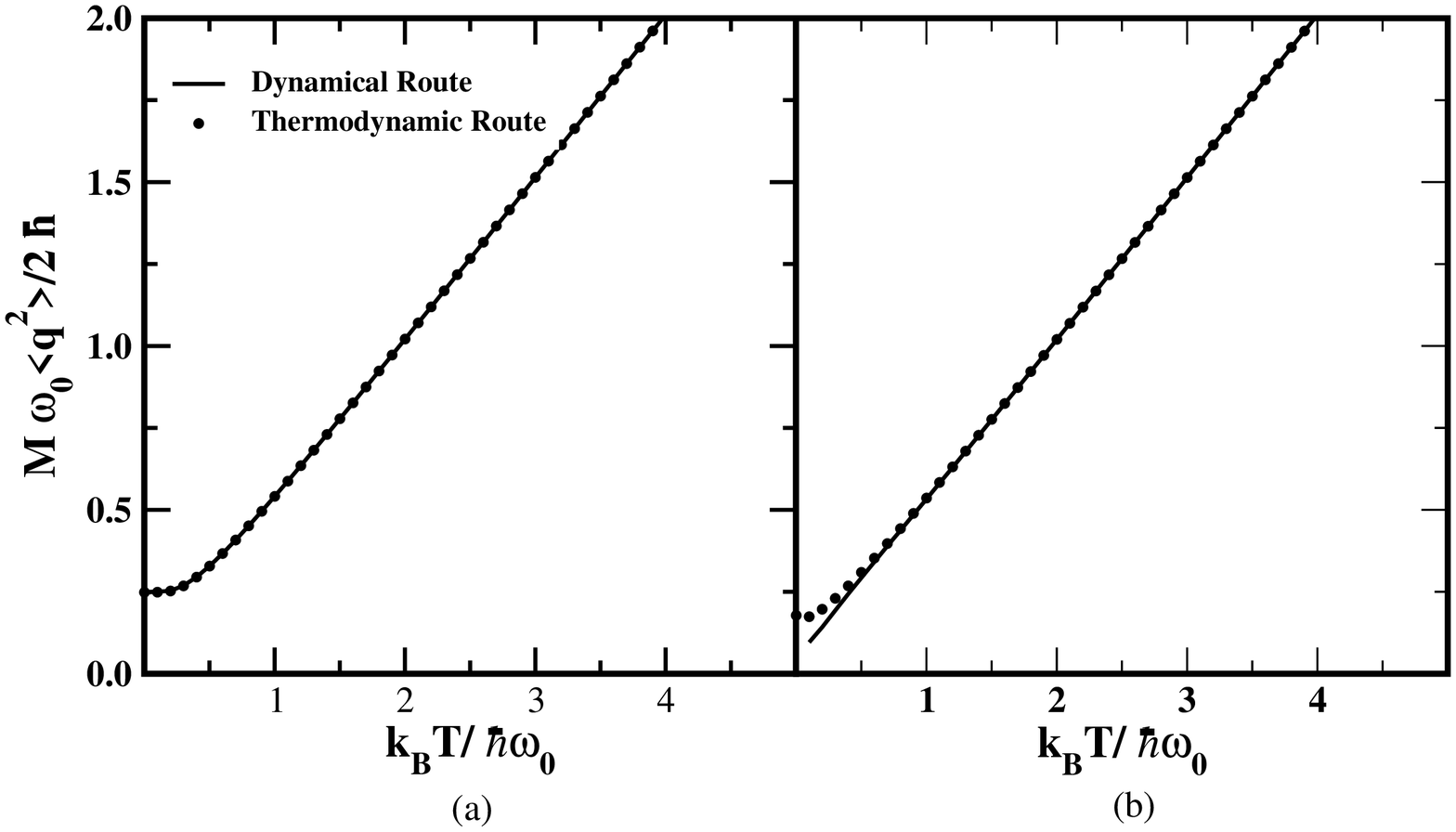}
\caption{The average potential energy  as a function of the temperature for (a) $\gamma/\omega_0$ = 0.01 and (b) $\gamma/\omega_0 = 2$. The solid line is obtained from Eq. \ref{energies} and the circles denote the thermodynamic result taken from \cite{weiss}}
\label{qsq}
\end{figure}
From the master equation (Eq. \ref{master-eq}), one can also derive the average kinetic and potential energies as the system approaches equilibrium\cite{positivity1}. The time derivative of the average values $q^2$, $p^2$ and $qp+pq$ can be calculated as,
\begin{widetext}
\begin{subequations}
\begin{align}
 \frac{d}{dt}\langle q^2 \rangle &=\frac{1}{M} \langle qp+pq \rangle +2D_{qq} \label{ddtq2} \\ 
\frac{d}{dt}\langle p^2 \rangle &= -M\omega_{0}^2 \langle qp+pq \rangle -4\gamma \langle p^2 \rangle +2D_{pp} \label{ddtp2} \\ 
\frac{d}{dt}\langle qp+pq \rangle &= \frac{2}{M}\langle p^2 \rangle -2M\omega_{0}^2 \langle q^2 \rangle -2\gamma \langle qp+pq \rangle -4D_{pq} \label{ddtqp}
\end{align}
\label{q2p2xp}
\end{subequations}
The solutions of the above coupled equation can then be obtained as,
\begin{subequations}
\begin{align}
 \langle q^2 \rangle &=\frac{(D_{pp}-4M\gamma D_{pq}+M^2 (4\gamma^2 +\omega_{0}^{2}) D_{qq})}{2M^2 \gamma \omega_{0}^2}  \nonumber \\  
&- C_1 \frac{e^{-2\gamma t}}{2M\gamma}-C_2 \frac{\gamma (\gamma +\Omega)e^{-2(\gamma-\Omega)t}}{2M\gamma \omega_{0}^2}
- C_3 \frac{\gamma(\gamma-\Omega)e^{-2(\gamma+\Omega)t}}{2M\gamma\omega_{0}^2} \label{q2av1} \\  
\langle p^2 \rangle &=\frac{(D_{pp}+M^2 \omega_{0}^2 D_{qq})}{2\gamma}-C_1 \frac{M\omega_{0}^2}{2\gamma}e^{-2\gamma t} \nonumber \\ 
&-C_2 \frac{M}{2\gamma}e^{-2(\gamma-\Omega)t}\gamma(\gamma-\Omega)- C_3 \frac{M}{2\gamma}e^{-2(\gamma+\Omega)t}\gamma(\gamma+\Omega) \label{p2av1} \\ 
\langle qp+pq \rangle &=-2MD_{qq}+C_1 e^{-2\gamma t}
+ C_2 e^{-2(\gamma-\Omega)t}+ C_3 e^{-2(\gamma+\Omega)t} \label{qpav1}
\end{align}
\label{averages-time}
\end{subequations}
\end{widetext}
Where $C_i$'s are the constants of integration, which depend on the initial values of the average quantities, viz. $\langle q^2 \rangle_0$, $\langle p^2 \rangle_0$ and $\langle qp+pq \rangle_0$ and,$\Omega=\sqrt{\gamma^2-\omega_{0}^2}$.
For finite oscillator frequency $\omega_0$, the equilibrium values of the average kinetic and potential energies can be obtained from,
\begin{subequations}
\begin{align}
\langle p^2 \rangle &=\frac{(D_{pp}+M^2 \omega_{0}^2 D_{qq})}{2\gamma}\label{p2av2} \\ 
\langle q^2 \rangle &=\frac{(D_{pp}-4M\gamma D_{pq}+M^2 (4\gamma^2 +\omega_{0}^{2}) D_{qq})}{2M^2 \gamma \omega_{0}^2} \label{q2av2}
\end{align}
\label{energies}
\end{subequations}

\begin{figure}[htb]
\vspace{0.5cm}
\begin{center}
\includegraphics[width=8cm]{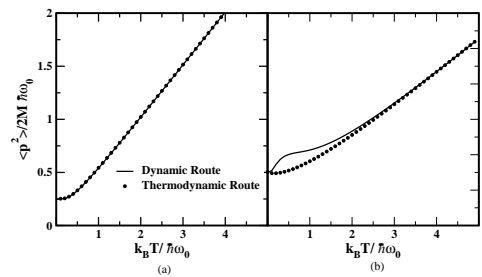}
\end{center}
\vspace{-0.1cm}
\caption{The average kinetic energy  as a function of the temperature for (a) $\gamma/\omega_0$ = 0.01 and (b) $\gamma/\omega_0 = 2$. The solid line is obtained from Eq. \ref{energies} and the circles denote the thermodynamic result taken from \cite{weiss}.}
\label{psq}
\end{figure}
At high temperatures, the average value of both kinetic energy $\langle p^2 \rangle/2M$ and potential energy
$M \omega_{0}^{2} \langle q^2 \rangle/2$ approach to $\frac{1}{2}k_{B} T$, which is in agreement with 
equipartition theorem. Thermodynamic quantities of the system can also be calculated from the partition function,
which can be evaluated by well known technique of imaginary time path integral method\cite{weiss,grabert}.
Above the critical temperature (below which the positivity condition is violated), equilibrium values of 
$\langle p^2 \rangle$ and $\langle q^2 \rangle$ obtained from the master equation are in good agreement
with those obtained from the partition function method as shown in Figs. \ref{qsq} and \ref{psq}.

\section{Revisiting the free particle : The limit of vanishing oscillator frequency}

It is obvious from the Hamiltonian Eq. \ref{hamil}, that, if we take $\omega_0 \rightarrow 0$, we would end up  with the system of a free particle connected to a heat bath. This problem was discussed in details by us in an earlier paper. We will present here the relevant results for the diffusion constants and the averages of $q^2$ and $p^2$ from the earlier paper. The diffusion constants are given by\cite{ss},
\begin{subequations}
\begin{align}
D_{qq} &=  \frac{2 k_B T \gamma \alpha^\prime_0}{M} \label{dqq-free}\\ 
D_{pq} &=  4 \gamma^2 k_B T \alpha^\prime_0 \label{dpq-free}\\ 
D_{pp} &=  2 k_B T M \gamma (1 + 4 \gamma^2 \alpha^\prime_0) \label{dpp-free}
\end{align}
\label{diffusion-free}
\end{subequations}
where, 
\begin{equation}
\alpha^\prime_0 = \frac{\frac{\hbar \gamma}{k_{B} T} \coth(\frac{\hbar \gamma}
{k_{B} T}) - 1}{4 \gamma^{2}},
\label{alpha_0_def}
\end{equation}
 and the corresponding averages for $q^2$ and $p^2$ are given by,
\begin{eqnarray}
\langle p^2 \rangle &=& \frac{D_{pp}}{2\gamma}  \nonumber \\
&=& M \hbar \gamma\coth\left(\frac{\hbar \gamma}{k_B T}\right)
\label{pav-free}
\end{eqnarray}
and,
\begin{eqnarray}
\langle q^2 \rangle &\sim& \frac{D_{pp}+4 M^2 \gamma^2 D_{qq}-4 M \gamma D_{pq}}{2 M^2 \gamma^2}t \nonumber \\
&=& \frac{k_B T}{M \gamma} t
\label{qav-free} 
\end{eqnarray}
From Eq. \ref{diff-const-1}, putting $\omega_0 \rightarrow 0$ we find that we can recover Eq. \ref{diffusion-free}, 
because in this limit $\alpha \rightarrow 1$ and $\alpha^\prime \rightarrow \alpha^\prime_0$. However, it is puzzling to note that Eq. \ref{qav-free} cannot be recovered from Eq. \ref{q2av2}, by taking the limit of $\omega_0 \rightarrow 0$. In order to recover the well known diffusive behaviour of the free particle, we must consider the system as it approaches equilibrium, viz. Eq. \ref{averages-time}, instead of the equilibrium values given by Eq. \ref{energies}. In Eq. \ref{averages-time}, as $\omega_0 \rightarrow 0$, one of the exponents ($\gamma - \Omega$) which sets the relaxation time vanishes. Thus, we expand this term as a power series, i.e.,$\exp(-2(\gamma-\Omega)t)$ $\sim$ $1+(\omega_0^2/2\gamma)t + {\cal O}(\omega_0^4)$. This term comes with the integration cosntant $C_2$, which can be calculated as,
\begin{widetext}
\begin{eqnarray}
C_2&=& \frac{(D_{pp}\Omega +[2m^2 \gamma^3 -2m^2 \gamma \omega_{0}^2 +2m^2 \gamma^2 \Omega -m^2 \Omega \omega_{0}^2]D_{qq}-2mD_{pq} [\Omega^2 +\gamma \Omega])}{2m\Omega^3} \nonumber \\
&+& \frac{(\langle p^2 \rangle _{t=0} [\Omega^2 -\gamma \Omega]-\langle x^2 \rangle _{t=0} m^2 \omega_{0}^2 [\Omega^2 +\gamma \Omega]-\langle qp+pq \rangle _{t=0} m\omega_{0}^2 \Omega)}{2m\Omega^3} 
\label{c2}
\end{eqnarray}
\end{widetext}
Inserting the power series expansion and the value of $C_2$ from Eq. \ref{c2}, in Eq. \ref{averages-time}, we recover Eqs. \ref{pav-free} and \ref{qav-free} exactly, in the limit of $\omega_0 \rightarrow 0$. Thus, it is clear that the equilibrium properties of a particle in a harmonic oscillator has an extra degree of freedom and hence the free particle results cannot be recovered in a straightforward manner.
\section{Conclusions}
We have considered a microscopic model of a particle in a harmonic well, connected to a bath of oscillators and derived a master equation for the reduced density matrix of the system. We have calculated the diffusion constants in both the overdamped and the underdamped regimes and examined the positivity conditions for the reduced density matrix. The postivity condition is not satisfied below a breakdown temperature $T_c$, which depends on the dimensionless parameter $\omega_0/\gamma$. This is not very unexpected, since we neglected the transients for $t \gg \hbar/k_B T$. In the high temperature limit, we find that the diffusion constants become independent of the oscillator frequency (upto ${\cal O}$ $(1/T)$). The kinetic and potential energy obtained from the steady state solutions are in good agreement with the thermodynamic results. We have also recovered the free particle results, in the limit of vanishing oscillator frequency. To recover the correct diffusive behaviour of the free particle, it is important to consider the approach to equilibrium, rather than the asymptotic equilibrium values.

\end{document}